 \documentclass[twocolumn]{webofc}
\usepackage{xcolor}
\usepackage[varg]{txfonts}   % Web of Conferences font
\usepackage{hyperref}
\usepackage{url}
\usepackage[rightcaption]{sidecap}
\hypersetup{colorlinks=true,citecolor=blue,urlcolor=blue,linkcolor=blue}
\begin{document}
\title{Extracting Contact Forces in Cohesive Granular Ensembles}

\author{\firstname{Abrar} \lastname{Naseer}\inst{1}\fnsep\thanks{\email{abrarnaseer@iisc.ac.in}} \and
        \firstname{Karen E.} \lastname{Daniels}\inst{2}\fnsep\thanks{\email{kdaniel@ncsu.edu}} \and
        \firstname{Tejas G.} \lastname{Murthy}\inst{1}\fnsep\thanks{\email{tejas@iisc.ac.in}}
        % etc.
}

\institute{Indian Institute of Science, Bangalore, India
\and
           Department of Physics, North Carolina State University, Raleigh, NC, USA
          }

\abstract{Interparticle cohesion is prevalent in stored powders, geological formations, and infrastructure engineering, yet a comprehensive understanding of the effects of its micro-mechanics on bulk properties has not been established experimentally.                                     %[needs some sort of question you'll answer: in this case HOW to address the idealized case needs solving, before other work can be done] 
One challenge has been that while photoelasticy has been widely and successfully used to measure the vector contact forces within dry granular systems, where the particle-particle interactions are solely frictional and compressive in nature, it has seen little development in systems where tensile forces are present.
%The existing photo-elasticity technique maps particle intensity-field to stress-field in dry granular materials, providing vector contact force information for every photo-elastic particle. On the contrary, in cohesive granular materials, particles can have tensile force interactions in addition to frictional and compressive interactions.
The key difficulty has been  the inability to distinguish between compressive and tensile forces, which appear identically within the photoelastic response. 

\hspace{5mm} Here, we present a novel approach which solves this problem, by an extension to the open-source PeGS (Photoelastic Grain Solver) software available at \url{https://github.com/photoelasticity}. Our new implementation divides the procedure of finding vector contact forces into two steps: first evaluating the vector contact forces on the non-cohesive particles present in the ensemble, followed by using an equilibrium constraint to solve for the forces in the bonded particles. We find that in the dilute limit, for up to 25\% bonded dimers, we can solve for all forces since each particle has only one force bearing contact that can potentially transmit tensile forces. %Using this new technique, we illustrate  that the inclusion of cohesion alters the force chain propagation and has a significant effect on the overall stiffness of the ensemble,  traceable to the presence of tensile forces.
While the case of dimers is an idealised version of cohesive granular ensemble, it provides an important first step towards experimentally studying the micro-mechanics of cohesive granular materials. %Our framework can be extended to more complex cohesive granular ensembles such as those involving linear chains of bonded particles. 
}
\maketitle
\section{Introduction}
\label{intro}
Photoelasticity has found  wide use in the study of granular materials. After pioneering work by Wakabayashi \cite{wakabayashi1950photo}, Majmudar \& Behringer \cite{majmudar2005contact} solved the inverse problem to obtain vector contact forces using photoelasticity, paving  the way for numerous quantitative, micro-mechanical studies on granular materials.
This quantification of vector contact forces has led to exploration of chute flows \cite{thomas2019photoelastic}, the yielding transition \cite{shang2024yielding}, shape effects \cite{wang2008characterization,wang2021contact}, force responses under varied boundary conditions \cite{zhang2014force}, rigidity percolation \cite{liu_spongelike_2021},  and shear jamming \cite{zhao2019shear} in granular materials. However, for many years the use of photoelastic techniques \cite{daniels2017photoelastic,abed2019enlightening} has focused  on the properties and behaviour of non-cohesive granular materials, leaving cohesive granular materials relatively unexplored. 

Yet, 
cohesion \cite{sharma2025experimental} is a widely encountered phenomenon in granular materials. In the presence of cohesion, the resistance of the granular material to shear and tensile forces improves drastically \cite{yang2018experimental,Amini2013ShearSC,Dass1994TensileSC}. The effect of cohesion on the granular materials has been well studied using ensemble-level experiments \cite{yang2018experimental,li2015experimental} as well as the discrete element method (DEM) simulations \cite{wang2008characterization,estrada2010simulation}. %\textcolor{red}{Although the bonds comprise pairwise interactions but there is a possibility that they additionally influence the surrounding particles in the cohesive ensemble. This many-body interaction \cite{hohler2017many} has not been well established in granular materials yet and requires particle-level study under controlled experiments}.
~To extend the use of photoelasticity to cohesive granular materials, we put forward an approach which extends an existing, open source method \cite{PeGS} of obtaining vector contact forces in granular materials to solve for dilute tensile forces. 
After describing how cohesion complicates the contact force extraction process, we present a method for overcoming these challenges, demonstrate that it works for dilute cohesion, and illustrate its effects on system-wide force transmission.
%The structure of the paper is as follows: in section \ref{sec:background} we demonstrate how introduction of cohesion complicates the contact force extraction process, followed by section \ref{sec-2} in which we present our novel approach to overcome this problem and therefore, solve for the tensile forces arising due to cohesion under the existing framework of photoelasticity.

\begin{SCfigure*}
\includegraphics[width=11cm,clip]{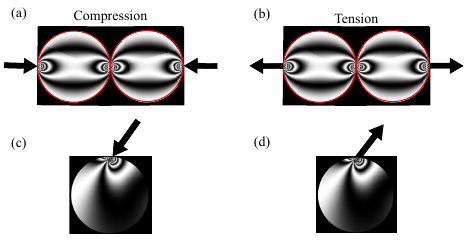}
\caption{Numerically solutions comparing the resulting fringe pattern for compression (left) and tension (right) under the inversion $f_n = -f_n$. Top row: diametric loading (no tangential forces). Bottom rows: the contribution from a single contact with the same tangential force.}
\label{fig:2examples}       % Give a unique label
\end{SCfigure*}

\section{Photoelastic methods \label{sec:background}}

The current implementation of photoelasticity \cite{ramesh, photoelasticity} in non-cohesive granular materials is well established for 2D planar systems and derives the vector contact forces  \cite{daniels2017photoelastic, abed2019enlightening} under the assumption that only compressive (non-tensile) forces are present, but those arising due to cohesion remain unexplored.
%Under this procedure, the nature of contact forces is restricted to being compressive - which is the mode of transmission in non-cohesive granular materials. 
When tensile forces are present, as is the case for cohesive granular materials, this creates an ambiguity in the method since normal contact forces $f_n$  and $-f_n$ create identical photoelastic responses. 
%The experimental evaluation of cohesive granular materials using photoelasticity is restricted due to the complexity in differentiating the fringe patterns under compression and tension. 
This is illustrated in Figure \ref{fig:2examples}, which compares the expected fringe pattern formed in a bonded pair of particles under diametric compressive and tensile loading.  
%In (a) under purely compressive loading, the point of contact/bond between the particles transmits compressive force. 
This arises because the stress field inside the particles results in the fringe pattern governed by 
%Eq. \ref{eq:1} - as can be seen in Fig. \ref{fig:2examples}(a)
\begin{equation}
    I(x,y) = I_0 \sin^2 \left( \frac{\pi (p - q) \, h \, C}{\lambda} \right)
    \label{eq:sin2}
\end{equation}
where $I_o$ is the overall light intensity, ($p$, $q$) are the principal stresses, $h$ is the particle thickness, $C(\lambda)$ is the wavelength-dependent stress-optic coefficient and $\lambda$ is the wavelength of the polarised light.
Because $p(x,y)$ and $q(x,y)$ are identical under an inversion of the normal force, Eq.~\ref{eq:sin2} does not distinguish the compressive and tensile cases. 
%In (b), similar external loading is applied but the nature of external force is flipped to tensile. Hence, the force transmitted through the bond changes to tensile. Due to the nature of mapping between the stress field and the intensity field, the fringe pattern is unchanged under the change in the nature of contact forces. 

Prior work has focused on dry granular materials, for which the PeGS algorithm uses an inverse mapping and optimization procedure which always selects the compressive force instead of the tensile force when solving for the normal force. The sensitivity to parameter choices has been well-documented \cite{liu_spongelike_2021,mcmillan2025validation} for that case, but 
%, \textit{i.e} convergence process works by relating the experimentally obtained fringe pattern to the best possible combination of vector contact forces, it therefore can not distinguish between a compressive and a tensile contact force as they both result in a similar fringe pattern. 
since the possibility of tensile forces creates an indeterminacy in the sign of the normal force which cannot be determined by this inverse process, additional information is required to select the correct solution. In general, forces are not colinear, as illustrated for a single frictionally-loaded contact in Fig.~\ref{fig:2examples}, for which the frictional contact could be either compressive or tensile and yet have the same fringe pattern. As we will see below, establishing torque and force balance within a bonded system is able to resolve this indeterminacy.

%The complexity increases when the loading case is not as simple as a collinear set of contact force vectors. In that case, satisfying equilibrium in each bonded particle adds to the complexity of the problem. This simple case demonstrates that the extension of established methodology to evaluate the vector contact forces in cohesive granular materials is restricted due to its inability to distinguish between a compressive and a tensile force, making the experimental evaluation of cohesive granular materials restricted. Therefore, we need to improve the current method with tensile force calculating capability to solve for the contact forces resisted by the bonded particles.
\begin{figure}[b]
\centering
\includegraphics[width=8cm,clip]{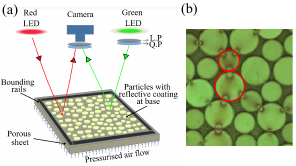}
\caption{(a) Schematic of the experimental setup: the red LED is used for particle detection, while a polarized green LED is used to visualize the fringe pattern in the loaded ensemble. (b) A subsection of the experiment shows a cluster of photoelastic particles with a pair of particles (marked with red circles) bonded together.}
\label{fig:exper}       
\end{figure}

\section{Introducing cohesion \label{sec:cohesion}}
\begin{figure*}
\centering
\includegraphics[width=15cm,clip]{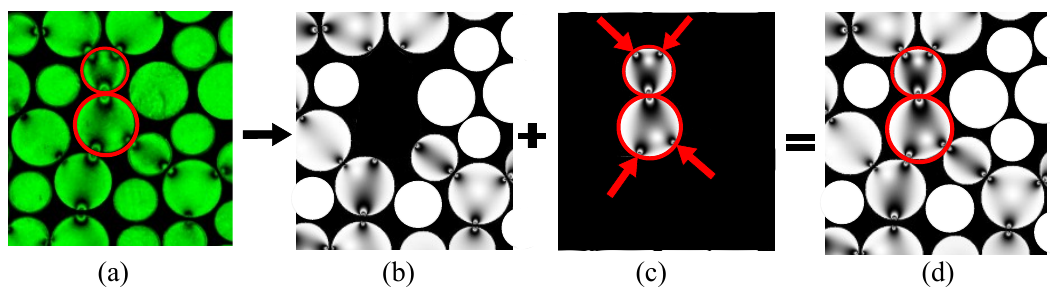}
\caption{
%Shows the steps involved in obtaining vector contact forces in a cohesive granular material. In (a) we first detect the bonded particles. In (b) we obtain the contact forces acting on all the non-bonded particles and therefore, reconstruct the pseudo image. In (c) contact forces are obtained for the isolated bonded pair of particles. (d)  The combined image from (b, c) gives the complete pseudo image of the experimental image.
Summary of the steps from an image to solved contact forces. Bonded particles (a) are initially removed from the analysis (b), after which they are solved using the forces (c) from their neighboring particles, to create the final result (d).}
\label{fig:process}      
\end{figure*}
We develop and test our new methods using experiments on a quasi-two-dimensional packing of photoelastic grains with controlled, dilute cohesion. Our particles rest above a horizontal porous sheet which floats them on a gentle layer of pressurised air (see Fig.~\ref{fig:exper}a), reducing the effects of basal friction on the force transmission. The particles are photoelastic (Vishay PSM-4) and are visualized within a brightfield polariscope. The combination of the linear polarizer (LP) and quarter plate (QP) placed in front of the green light source and the overhead camera is used to extract the force network by rotation of the polarisation of the green light. We introduce cohesion by randomly bonding pairs of particles with white glue at fixed locations which we identify in advance. To load the sample and create a force network, we provide a small amount of biaxial compression from two of the walls; a subregion of the granular material containing a bonded pair is shown in Figure~\ref{fig:exper}b.

To obtain the vector contact forces and hence a well-matched pseudo-image in a cohesive granular material, we divide the procedure into three steps. In the very first step, we locate the particles that are bonded in the ensemble by assigning them unique IDs for detection in the ensemble. In Fig.~\ref{fig:process}a, this is shown by two circles circumscribing the bonded pair. %This process is automated and therefore, the bonded particles are labelled in each strain step. 
For all particles except the bonded ones, the PeGS data analysis pipeline proceeds as usual, since none of these will contain tensile forces.
%Once the bonded particles are located, the particle pair is first removed from the Matlab \textit{structure} storing the particle information, and the remaining non-bonded particles are fed to the PeGS pipeline for analysis. 
Since PeGS evaluates vector contact forces particle-wise, this removal of bonded particles does not affect the convergence procedure for the non-bonded particles. This process is illustrated in the reconstructed pseudo image in Fig.~\ref{fig:process}b. Next, we evaluate forces on all the neighbouring particles of the bonded pairs:  by Newton's Third Law (reciprocity), this determines the external forces acting on each of the bonded particles, as shown in Fig.~\ref{fig:process}c. 
%on all the force-bearing contacts of the bonded pair (except the force transmitting through the bond). These forces can be seen as external forces on each particle forming the bonded pair. 
% Newton's third law of motion, we transfer the forces from neighbouring particles to the bonded pair (red arrows in Fig. \ref{fig:process}c). 
Finally, by invoking force and torque balance on the individual particles, we can solve for the contact force ($f_{x}$,$f_{y}$) transmitted through the bond itself.
%With two equations of equilibrium and two unknowns ($f_{x}$,$f_{y}$) we obtain all the vector contact forces acting on the bonded particles and
As shown in Fig.~\ref{fig:process}d, the reconstructed fringe pattern in the bonded pair matches well with its experimental image. It should be noted that while solving for the normal and tangential components, the force-bearing contacts follow a calibrated force model based on linear elasticity, as mentioned in Ref. \cite{daniels2017photoelastic,abed2019enlightening}.

\begin{figure}
\centering
\includegraphics[width=6cm,clip]{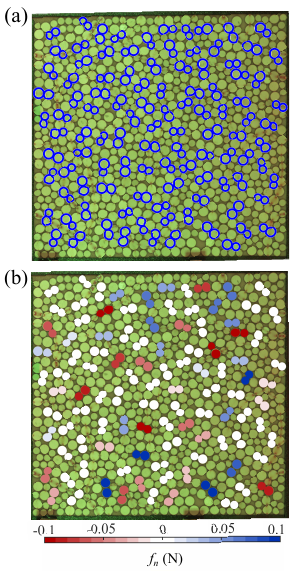}
\caption{Example of (a) a fully-solved granular packing with 25\% bonded pairs (shown by blue outlines), and the resulting force network and (b) the fluctuating nature of normal force ($f_n$) transmitting through the bonds. The coloured pairs are the bonded particles, where particles in the red spectrum bear tensile forces, and in the blue are compressed.}
\label{fig:wholesystem}       
\end{figure}

Therefore, we find that by separating the process of extracting the vector contact forces into two steps --- first, analyzing the non-bonded particles, and second, handling the bonded particles --- we can reconstruct a combined image that accurately represents the complete pseudo-image of the experimental image. Note that this result neglects the smallest forces, for which the fringe pattern may have an intensity below the threshold used to detect force-bearing contacts. This process has been successful in processing granular materials with up to 25\% bonded pairs \cite{naseer_preprint}. An example is shown in Fig.~\ref{fig:wholesystem} where the entire cohesive ensemble is biaxially compressed, resulting in the formation of force network. In addition to
compressive forces, the vector force data reveals the presence of tensile interactions at the bonds (Fig.~\ref{fig:wholesystem}b). Particle pairs shown in red colour are the ones experiencing net tensile forces through their interparticle bonds, whereas
those in blue colour experience net compressive forces. This finding opens the regime of comprehensive experimental study of the particle-level features of cohesive granular ensembles.
 
%Importantly, in this approach, the nature of the force transmitted through the bond is now well understood.

\section{Conclusion}
Our new implementation \cite{naseer_cohesion}, soon to be included in the new release of  PeGS \cite{PeGSv2.0},  allows us to extend photoelastic techniques to work on cohesive granular materials. This procedure will likely also apply to systems where particles are bonded in longer linear structures, or even to larger clusters by starting from the edges (only one bond/particle) and working inwards iteratively.
%The ensemble needs to be carefully constructed to avoid any bonded pair of particles (or chains) coming in contact with each other - leading to the indeterminacy of the system.
Our model bonds closely resemble any bond that has a rigidity similar to the particle rigidity, such as cementation, solid bridges, or a polymer coating. %underlying principle of constraining the degrees of freedom by bonding remains consistent.}  
Our improvement will enable future work on the micro-mechanics of cohesive granular materials.

\section{Acknowledgments}
We are grateful to funding from the National Science Foundation (DMR-2104986), Anusandhan National Research Foundation (ANRF-CRG/2022/003750)  and the Fulbright-Nehru fellowship program.
\bibliographystyle{unsrt}
 \bibliography{bib_PnG} 
\end{document}